\documentclass[journal,twocolumn]{IEEEtran}   
\usepackage{graphicx}
\usepackage{epstopdf}
\usepackage{amssymb}
\usepackage{amsfonts}
\usepackage{amsmath}
\usepackage{booktabs}
\usepackage{algorithm}
\usepackage{algorithmic}
\usepackage{subeqnarray}
\usepackage{cases}
\usepackage{amsmath,amsfonts}
\usepackage{bm}
\usepackage{subfigure,amsmath,amssymb,cite}
\usepackage{stfloats}
\usepackage{booktabs}

\usepackage{float}
\ifCLASSINFOpdf
\else
\fi
\begin{document}

\title{Enhanced Secure Wireless Transmission Using IRS-aided Directional Modulation}

\author{Yeqing Lin, Rongen Dong,~Peng Zhang,~Feng Shu,~and Jiangzhou Wang,\emph{ Fellow, IEEE}

\thanks{This work was supported in part by the National Natural Science Foundation of China (Nos. 62071234, 62071289, and 61972093), the Hainan Province Science and Technology Special Fund (ZDKJ2021022), the Scientific Research Fund Project of Hainan University under Grant KYQD(ZR)-21008, and the National Key R\&D Program of China under Grant 2018YFB180110.(Corresponding authors: Feng Shu and Rongen Dong)}
\thanks{Yeqing Lin, Rongen Dong, Peng Zhang, and Feng Shu are with the School of Information and Communication Engineering, Hainan University, Haikou, 570228, China. (Email: shufeng0101@163.com) }
\thanks{Feng Shu is with the School of Electronic and Optical Engineering, Nanjing University of Science and Technology, 210094, China. (Email: shufeng0101@163.com)}
\thanks{Jiangzhou Wang is with the School of Engineering, University of Kent, Canterbury CT2 7NT, U.K. (Email: j.z.wang@kent.ac.uk)}


}

\maketitle

\begin{abstract}
As an excellent aided communication tool, intelligent reflecting surface (IRS) can make a significant rate enhancement and  coverage extension.  In this paper, we present an investigation on beamforming in an IRS-aided directional modulation (DM) network. To fully explore the advantages of IRS, two beamforming methods with enhanced secrecy rate (SR) performance are proposed. The first method of maximizing secrecy  rate (Max-SR) alternately optimizes confidential message (CM) beamforming vector, artificial noise (AN) beamforming vector and phase shift matrix. The first optimization vector is directly computed by the Rayleigh ratio and the last two are solved with generalized power iteration (GPI). This method is called Max-SR-GPI. To reduce the computational complexity, a new method of maximizing receive power with zero-forcing constraint (Max-RP-ZFC) of only reflecting CM and no AN is proposed. Simulation results show that the proposed two methods harvest about 30 percent rate gains over the cases of random-phase IRS and no IRS, and the proposed Max-SR-GPI performs slightly better than the Max-RP-ZFC in terms of SR, particularly in the small-large IRS.
\end{abstract}
\section{Introduction}
With the development of the sixth generation of mobile communications, intelligent reflecting surface (IRS) has been promised as a key technology to enhance rate, extend coverage and remove blind areas \cite{Wuqingqing2020,Pangxiaowei2022,Wuqingqing2019,Chengxin2021}. IRS is becoming increasingly important in such diverse communication areas as multiple input and multiple output (MIMO) \cite{Pancunhua2020}, spatial and directional modulation networks \cite{Jiangxinyi2022}, relay \cite{Wangxuehui2022}, and covert \cite{Zhouxiaobo2022}.
To utilize IRS to increase the achievable rate of the downlink, the authors proposed the joint optimization of transmitter beamforming, IRS phase shift, IRS orientation, and position in IRS-assisted multiple-input-single-output (MISO) free-space wireless transmission systems in \cite{Chengxin2021}. In \cite{Pancunhua2020}, the authors confirmed that the cell-edge user performance can be improved by using IRS in the case of downlink multi-user MISO.
In\cite{Jiangxinyi2022}, IRS was proposed to assist spatial modulation that maximizes secrecy rate (SR) by adjusting the switching state of  the IRS reflecting elements by power control. However, the transmission behavior of the transmitter, once detected by a malicious node, exposed the network to a security risk, in \cite{Zhouxiaobo2022}, the authors designed IRS-assisted and AN-enhanced wireless covert communication to achieve covert transmission rate multiplication, and more importantly proved the existence of perfect covertness under perfect channel state information (CSI). In \cite{Wanghuiming2020}, considering a more realistic scenario without Eve's CSI, which jointly optimized beamforming and interference to satisfy the quality of service for Bob and emitted artificial noise (AN) to interfere with Eve. In \cite{Guanxinrong2020}, the authors demonstrated that using AN can be an effective way to help improve the security. In addition to using one IRS, the authors used two or more IRSs to further enhance the system performance in \cite{Hanyitao2022,Yangliang2022}.

As an advanced physical layer security technique, directional modulation (DM) was well suited for line-of-sight channel and implemented secure precise wireless transmission with the help of  AN, random subcarrier selection, and beamforming in \cite{Wuxiaoming2016,Binqiu2019}.  In \cite{Lijiayu2021}, with the aid of IRS, DM can achieve  two-way independent CM streams from Alice to Bob in multipath channel, here IRS may control the phases of path gains. However, the proposed two methods required high computational amounts. In \cite{Hujingsong2020}, a single CM symbol was transmitted from Alice to Bob using two symbol periods, which results in a significant SR loss. In this paper, we still focus on a single CM stream transmission with transmitting one CM symbol per symbol period. Two low-complexity methods are proposed to strike a good balance between complexity and performance. The main contributions of this paper are summarized as follows:
\begin{enumerate}
  \item A system of combining IRS and DM is established to realize an enhanced single CM stream by fulling making use of the advantages of DM and IRS. To achieve an improved SR performance, a method of maximizing the SR is proposed to alternately optimize the CM beamforming vector, AN beamforming vector, and IRS phase shift matrix. The first optimization vector (OV) is computed by the Rayleigh ratio, and the last two OVs are solved by converting their optimization problem into the GPI canonical forms. Since the OVs of the latter two are solved by the GPI algorithm, and the optimal SR obtained by this method is related to the initial value. Therefore, this method has high complexity.
  \item To reduce the high complexity of the above method, the new method is proposed as follows:~IRS only reflects CM and no AN. In other words, the AN beamforming  vector is orthogonal to both channels from Alice to IRS and from Alice to Bob. Under the zero-forcing constraint (ZFC),  maximizing the receive power (Max-RP) is proposed to compute the CM beamforming vector, IRS phase matrix and AN beamforming vector, respectively. Hence, this method is called Max-RP-ZFC. Compared with Max-SR-GPI method, the proposed Max-RP-ZFC method has lower computational complexity. However, the latter performs slightly worse than the former in the case of small-scale IRS. As the number of IRS elements tends to large-scale, the SR difference between them becomes trivial.
\end{enumerate}

The remainder is organized as follows. Section II presents the system model and two methods are proposed in Section III. In Section IV, numerical simulations are presented, and Section V draws our conclusion.

$Notations$: In this paper, bold lowercase and uppercase letters represent vectors and matrices, respectively.
{Signs  $(\cdot)^H$, $(\cdot)^{-1}$, $\operatorname{tr}(\cdot)$, and $\|\cdot\|$ denote the conjugate transpose operation, inverse operation, trace operation, and 2-norm operation, respectively. The notation $\textbf{I}_N$ is the $N\times N$ identity matrix. The sign $\mathbb{E}\{\cdot\}$ represents the expectation operation, and $\operatorname{diag}(\cdot)$ denotes the diagonal operator.
\section{system model}
\begin{figure}[h]
\centering
\includegraphics[width=0.45\textwidth,height=0.23\textheight]{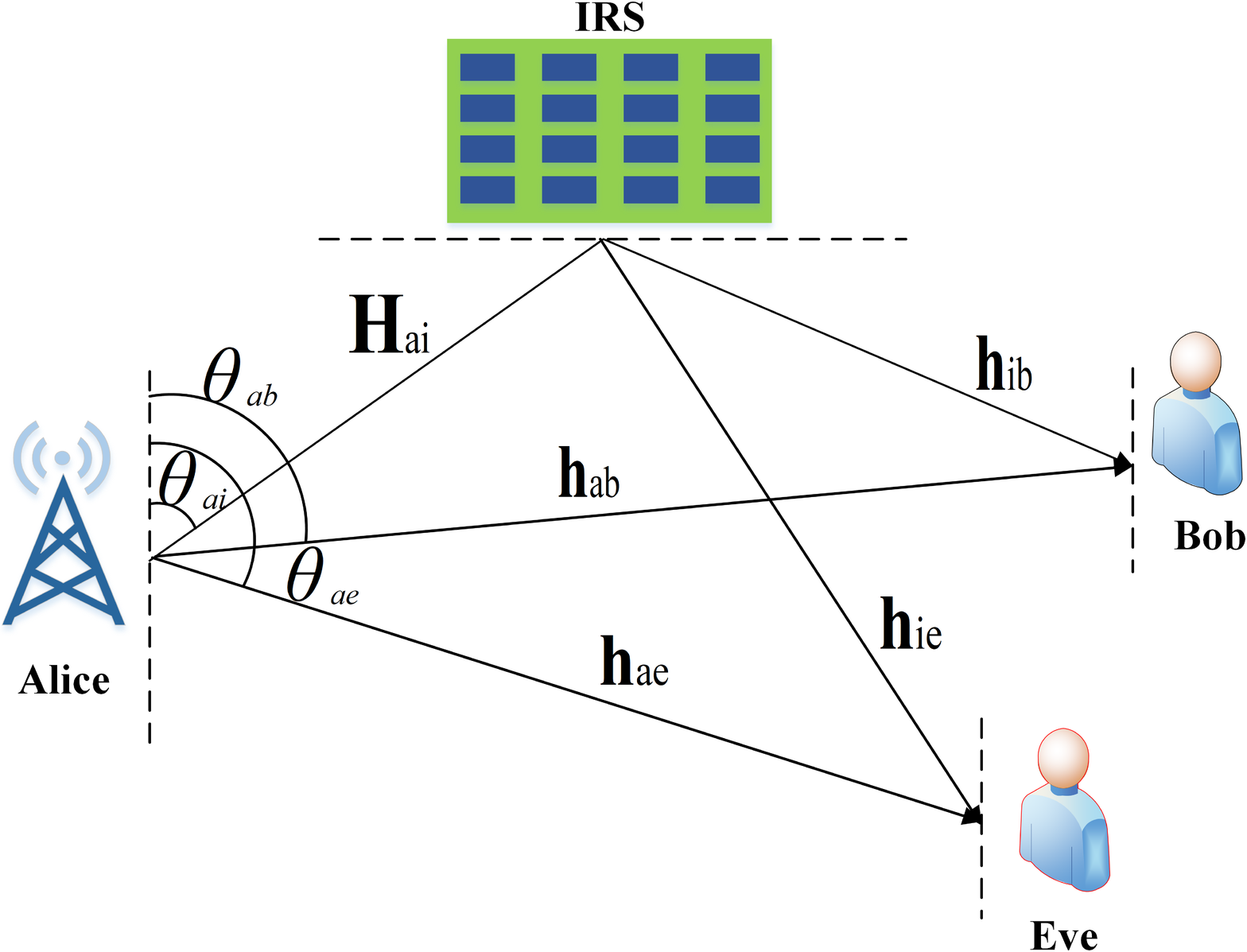}\\
\caption{System model diagram of IRS-aided DM.}\label{systemmodel.eps}
\end{figure}
In Fig. \ref{systemmodel.eps}, an IRS-aided DM network is presented. Alice is equipped with $N_{a}$  antennas, IRS has $N_{r}$  reflecting elements, and Bob and Eve are employed with single antenna.

The transmit baseband signal is in the form
\begin{equation}
\mathbf{x}=\sqrt{\beta_{1} P_{t}}\mathbf{v}_{a} s_{a}+\sqrt{\beta_{2} P_{t}} \mathbf{v}_{A N}{z},
\end{equation}
where $P_{t}$ is the transmit power, $s_{a}$ denotes the CM  with a constraint $\mathbb{E}\left[|s_{a}|^{2}\right]=1$, $z$ is the AN with the average power constraint $\mathbb{E}\left[|z|^{2}\right]=1$, $\mathbf{v}_{a}$ denotes the precoding vector of CM,  and $\mathbf{v}_{A N}$ is the precoding vector of AN with $\mathbf{v}_{a} \in \mathbb{C}^{N_{a} \times 1}$  and $\mathbf{v}_{AN} \in \mathbb{C}^{N_{a} \times 1}$.  $\beta_{1}$ and $\beta_{2}$  respectively represent the power allocation (PA) factors of CM and AN with $\beta_{1}+\beta_{2}=1$.
The signal received at Bob can be represented as
\begin{align}\label{y_b}
&y_{b}=\left(\sqrt{g_{a b}} \mathbf{h}_{a b}^{H}+\sqrt{g_{a i b}} \mathbf{h}_{i b}^{H} \bm{\Theta} \mathbf{H}_{ai}\right)\mathbf{x}+n_{b}\nonumber\\
=&\sqrt{g_{a b} \beta_{1} P_{t}} \mathbf{h}_{a b}^{H} \mathbf{v}_{a} s_{a}+\sqrt{g_{a i b} \beta_{1} P_{t}} \mathbf{h}_{i b}^{H} \bm{\Theta} \mathbf{H}_{ai}\mathbf{v}_{a} s_{a}+n_{b} \nonumber\\
&\sqrt{g_{a b} \beta_{2} P_{t}} \mathbf{h}_{a b}^{H} \mathbf{v}_{AN} z+\sqrt{g_{a i b} \beta_{2} P_{t}} \mathbf{h}_{i b}^{H} \bm{\Theta} \mathbf{H}_{ai}\mathbf{v}_{AN} z,
\end{align}
where $g_{ab}$ is the path loss coefficient between Alice and Bob, $g_{aib}$=$g_{ai}g_{ib}$ denotes the equivalent path loss coefficient of Alice-to-IRS channel and IRS-to-Bob channel, $\mathbf{h}_{ab} \in \mathbb{C}^{N_{a} \times 1}$ represents the Alice-to-Bob channel, $\mathbf{h}_{ib}$ $\in \mathbb{C}^{{N_{r}} \times 1}$ represents the IRS-to-Bob channel, $\boldsymbol{\Theta}=\operatorname{diag}\left(e^{j \phi_{1}}, \cdots, e^{j \phi_{m}}, \cdots, e^{j \phi_{N_{r}}}\right)$ is a diagonal matrix with the phase shift $\phi_{m}$ incurred by the $m$-th reflecting element of the IRS, $\bm{\Theta}=\operatorname{diag}(\bm{\theta})$ with $\bm{\theta}\in \mathbb{C}^{N_{r} \times 1}$, $\mathbf{H}_{ai}=\mathbf{h}\left(\theta_{ai,r}\right) \mathbf{h}^{H}\left(\theta_{ai,t}\right)\in \mathbb{C}^{N_{r} \times {N_{a}}}$ represents the Alice-to-IRS channel, and $n_{b} \sim \mathcal{C} \mathcal{N}\left(0, \sigma_{b}^{2}\right)$ is the additive white Gaussian noise (AWGN) at Bob.

Similarly, the signal received at Eve can be expressed as
\begin{align}\label{y_e}
&y_{e}=\left(\sqrt{g_{a e}} \mathbf{h}_{a e}^{H}+\sqrt{g_{a i e}} \mathbf{h}_{i e}^{H} \bm{\Theta} \mathbf{H}_{ai}\right)\mathbf{x}+n_{e}\nonumber\\
=&\sqrt{g_{a e} \beta_{1} P_{t}} \mathbf{h}_{a e}^{H} \mathbf{v}_{a} s_{a}+\sqrt{g_{a i e} \beta_{1} P_{t}} \mathbf{h}_{i e}^{H} \bm{\Theta} \mathbf{H}_{ai}\mathbf{v}_{a} s_{a}+n_{e} \nonumber\\
&\sqrt{g_{a e} \beta_{2} P_{t}} \mathbf{h}_{a e}^{H} \mathbf{v}_{AN} z+\sqrt{g_{a i e} \beta_{2} P_{t}} \mathbf{h}_{i e}^{H} \bm{\Theta} \mathbf{H}_{ai}\mathbf{v}_{AN} z,
\end{align}
where $g_{ae}$ is the path loss coefficient between Alice and Eve, $g_{aie}$=$g_{ai}g_{ie}$ denotes the equivalent path loss coefficient of Alice-to-IRS channel and IRS-to-Eve channel, $\mathbf{h}_{ae} \in \mathbb{C}^{N_{a} \times 1}$ represents the Alice-to-Eve channel, $\mathbf{h}_{ie}$ $\in \mathbb{C}^{{N_{r}} \times 1}$ represents the IRS-to-Eve channel, and $n_{e} \sim \mathcal{C} \mathcal{N}\left(0, \sigma_{e}^{2}\right)$  is the AWGN at Eve. The normalized steering vector $\mathbf{h}(\theta)$ is defined as
\begin{equation}
\mathbf{h}(\theta)=\frac{1}{\sqrt{N}}\left[e^{j 2 \pi \Phi_{\theta}\left(1\right)}, \ldots,e^{j 2 \pi \Phi_{\theta}\left(n\right)}, \ldots, e^{j 2 \pi \Phi_{\theta}\left(N\right)}\right]^{H},
\end{equation}
where the phase shift $\Phi_{n}(\theta)$ is defined as
\begin{align}\label{Phi_{n}}
\Phi_{n}(\theta)=-\frac{d}{\lambda}\left(n-\frac{N+1}{2}\right) \cos \theta, \quad n=1, \ldots, N,
\end{align}
where $\lambda$ is the wavelength, $n$ is the index of antenna, $d$ represents the element spacing in the transmit antenna array, and $\theta$ is the direction of departure.

The signal received at IRS can be expressed as
\begin{equation}\label{y_irs}
y_{i}=\mathbf{H}_{ai}\mathbf{x}
=\mathbf{H}_{ai}(\sqrt{\beta_{1} P_{t}}\mathbf{v}_{a} s_{a}+\sqrt{\beta_{2} P_{t}} \mathbf{v}_{A N}{z}).
\end{equation}
In terms of (\ref{y_b}), the signal-to-interference and noise ratio (SINR) at Bob is
\begin{equation}
\gamma_{b}=\frac{\beta_{1} P_{t}\left|\sqrt{g_{a b}} \mathbf{h}_{ab}^{H} \mathbf{v}_{a}+\sqrt{g_{a i b}} \mathbf{h}_{ib}^{H} \bm{\Theta} \mathbf{H}_{ai}\mathbf{v}_{a}\right|^{2}}{\beta_{2} P_{t}\left|\sqrt{g_{a b}} \mathbf{h}_{ab}^{H} \mathbf{v}_{A N}+\sqrt{g_{a i b} } \mathbf{h}_{ib}^{H} \bm{\Theta} \mathbf{H}_{ai} \mathbf{v}_{A N}\right|^{2}+\sigma_{b}^{2}}.
\end{equation}
From (\ref{y_e}), the SINR of Eve is
\begin{equation}
\gamma_{e}=\frac{\beta_{1} P_{t}\left|\sqrt{g_{a e}} \mathbf{h}_{ae}^{H} \mathbf{v}_{a}+\sqrt{g_{a i e}} \mathbf{h}_{ie}^{H} \bm{\Theta} \mathbf{H}_{ai}\mathbf{v}_{a} \right|^{2}}{\beta_{2} P_{t}\left|\sqrt{g_{ae}}\mathbf{h}_{ae}^{H} \mathbf{v}_{A N} +\sqrt{g_{a i e} } \mathbf{h}_{ie}^{H} \bm{\Theta} \mathbf{H}_{ai} \mathbf{v}_{AN}\right|^{2}+\sigma_{e}^{2}}.
\end{equation}
 The corresponding rates at Bob and Eve are as follows
 \begin{equation}
 R_{b}=\log _{2}\left(1+\gamma_{b}\right),
 \end{equation}
and
 \begin{align}
 R_{e}=\log _{2}\left(1+\gamma_{e}\right),
 \end{align}
 respectively, which directly give the secrecy rate as
\begin{align}\label{SR}
R_{s}=\left[R_{b}-R_{e}\right]^{+}=\log _{2}\left(\frac{1+\gamma_{b}}{1+\gamma_{e}}\right),
\end{align}
where $[x]^{+} \triangleq \max \{0, x\}$.

\section{Two proposed beamforming methods with enhanced performance}
In what follows, to harvest the SR performance gain available by IRS,  two iterative methods, called Max-SR-GPI and Max-RP-ZFC, are proposed. The former is  high-performance while the latter is low-complexity.
\subsection{Proposed Max-SR-GPI}
The optimization problem of maximizing the SR can be casted as
\begin{subequations}\label{Max-SR}
\begin{align}
&\max_{\mathbf{v}_{a}, \mathbf{v}_{AN}, \boldsymbol{\Theta}} R_{s}\left(\mathbf{v}_{a}, \mathbf{v}_{AN}, \boldsymbol{\Theta}\right) \\
&~~~\text { s.t. }~ \mathbf{v}_{a}^{H} \mathbf{v}_{a}=1, \mathbf{v}_{AN}^{H} \mathbf{v}_{AN}=1, \left|\Theta_{i}\right|=1.
\end{align}
\end{subequations}
 The rate of Bob in the above SR can be rewritten as
\begin{align}\label{Rb}
R_{b}=\log _{2}\left(1+\frac{\mathbf{v}_{a}^{H} \mathbf{h}_{B1}^{H} \mathbf{h}_{B1} \mathbf{v}_{a}}{\mathbf{v}_{AN}^{H} \mathbf{h}_{B2}^{H} \mathbf{h}_{B2} \mathbf{v}_{A N}+\sigma_{b}^{2}}\right),
\end{align}
where\begin{align}
\mathbf{h}_{B1}&=\left(\sqrt{\beta_{1} P_{t} g_{a b}} \mathbf{h}_{ab}^{H}+\sqrt{\beta_{1} P_{t} g_{a i b}} \mathbf{h}_{ib}^{H} \bm{\Theta} \mathbf{H}_{ai}\right),\nonumber\\
\mathbf{h}_{B2}&=\left(\sqrt{\beta_{2} P_{t} g_{a b}} \mathbf{h}_{ab}^{H}+\sqrt{\beta_{2} P_{t} g_{a i b}} \mathbf{h}_{ib}^{H} \bm{\Theta} \mathbf{H}_{ai}\right).
\end{align}
Similarly, the rate of Eve can be rewritten as
\begin{align}\label{Re}
R_{e}=\log _{2}\left(1+\frac{\mathbf{v}_{a}^{H} \mathbf{h}_{E1}^{H} \mathbf{h}_{E1} \mathbf{v}_{a}}{\mathbf{v}_{AN}^{H} \mathbf{h}_{E2}^{H} \mathbf{h}_{E2} \mathbf{v}_{A N}+\sigma_{e}^{2}}\right),
\end{align}
where
\begin{align}
\mathbf{h}_{E1}=\left(\sqrt{\beta_{1} P_{t} g_{a e}} \mathbf{h}_{ae}^{H}+\sqrt{\beta_{1} P_{t} g_{a i e}} \mathbf{h}_{ie}^{H} \bm{\Theta} \mathbf{H}_{ai}\right),\nonumber\\
\mathbf{h}_{E2}=\left(\sqrt{\beta_{2} P_{t} g_{a e}} \mathbf{h}_{ae}^{H}+\sqrt{\beta_{2} P_{t} g_{a i e}} \mathbf{h}_{ie}^{H} \bm{\Theta} \mathbf{H}_{ai}\right).
\end{align}
According to (\ref{Rb}) and (\ref{Re}), given $\bm{\Theta}$ and $\mathbf{v}_{AN}$, the optimization problem in (\ref{Max-SR}) is converted into
\begin{equation}
\begin{aligned}
&\max _{\mathbf{v}_{a}} \frac{\mathbf{v}_{a}^{H}\left((a+\sigma_{b}^2) \mathbf{I}_{N_{a}}+\mathbf{h}_{B1}^{H} \mathbf{h}_{B1}\right) \mathbf{v}_{a}}  {\mathbf{v}_{a}^{H}\left((b+\sigma_{e}^2) \mathbf{I}_{N_{a}}+\mathbf{h}_{E1}^{H} \mathbf{h}_{E1}\right) \mathbf{v}_{a}}
&\text { s.t. } &\quad \mathbf{v}_{a}^{H} \mathbf{v}_{a}=1,\\
\end{aligned}
\end{equation}
where $a=\mathbf{v}_{AN}^{H}\mathbf{h}_{B2}^{H}\mathbf{h}_{B2}\mathbf{v}_{AN}$, and $b=\mathbf{v}_{AN}^{H}\mathbf{h}_{E2}^{H}\mathbf{h}_{E2}\mathbf{v}_{AN}$ due to the fact that the logarithm function is a monotonically increasing function.

Therefore, using the Rayleigh-Ritz ratio theorem, $\mathbf{v}_{a}$ is the eigenvector corresponding to the largest eigenvalue of the following matrix
\begin{equation}\label{va}
{\left((b+\sigma_{e}^2) \mathbf{I}_{N_{a}}+\mathbf{h}_{E1}^{H} \mathbf{h}_{E1}\right)}^{-1}{\left((a+\sigma_{b}^2) \mathbf{I}_{N_{a}}+\mathbf{h}_{B1}^{H} \mathbf{h}_{B1}\right)}.
\end{equation}

 Similarly, given $\bm{\Theta}$ and $\mathbf{v}_{a}$, the optimization problem in (\ref{Max-SR}) is converted into
\begin{equation}\label{van}
\begin{aligned}
\max _{\mathbf{v}_{AN}} &\quad \frac{\mathbf{v}_{AN}^{H} \mathbf{E} \mathbf{v}_{AN}}{\mathbf{v}_{AN}^{H} \mathbf{F} \mathbf{v}_{AN}} \times \frac{\mathbf{v}_{AN}^{H} \mathbf{M} \mathbf{v}_{AN}}{\mathbf{v}_{AN}^{H} \mathbf{N} \mathbf{v}_{AN}} &\text{s.t.} &\quad \mathbf{v}_{AN}^{H} \mathbf{v}_{AN}=1,\\
\end{aligned}
\end{equation}
where
\begin{align}
&c=\mathbf{v}_{a}^{H}\mathbf{h}_{B1}^{H}\mathbf{h}_{B1}\mathbf{v}_{a}, d=\mathbf{v}_{a}^{H}\mathbf{h}_{E1}^{H}\mathbf{h}_{E1}\mathbf{v}_{a},\nonumber\\
&\mathbf{E}=(c+\sigma_{b}^2) \mathbf{I}_{N_{a}}+\mathbf{h}_{B2}^{H} \mathbf{h}_{B2}, \mathbf{F}=\sigma_{b}^2 \mathbf{I}_{N a}+\mathbf{h}_{B2}^{H}\mathbf{h}_{B2},\nonumber\\
&\mathbf{M}=\sigma_{e}^2\mathbf{I}_{N_{a}}+\mathbf{h}_{E2}^{H}\mathbf{h}_{E2}, \mathbf{N}=(d+\sigma_{e}^2)\mathbf{I}_{N_{a}}+\mathbf{h}_{E2}^{H} \mathbf{h}_{E2}.
\end{align}
Therefore, $\mathbf{v}_{AN}$ can be solved by using GPI algorithm in \cite{lee2008GPI}.

Given $\mathbf{v}_{AN}$ and $\mathbf{v}_{a}$, let us define a new optimization variable $\overline{\boldsymbol{\theta}}=\left[1,\boldsymbol{\theta}^{H}\right]^{H}$.
Accordingly, the rate of Bob can be rewritten as
\begin{align}
R_{b}=\log _{2}\left(1+\frac{P_{t} \left(\overline{\bm{\theta}}^{H} \mathbf{w}^{H} \mathbf{w} \overline{\bm{\theta}}\right)}{P_{t} \left(\overline{\bm{\theta}}^{H} \mathbf{v}^{H} \mathbf{v} \overline{\bm{\theta}}\right)+\sigma_{b}^{2}}\right),
\end{align}
where
\begin{align}
&\mathbf{w} = [\sqrt{\beta_{1}  g_{a b}} \mathbf{h}_{a b}^{H}  \mathbf{v}_{a} \quad \sqrt{\beta_{1}  g_{a i b}} \mathbf{h}_{i b}^{H}\operatorname{diag}\left(\mathbf{H}_{a i}\mathbf{v}_{a} \right)],\nonumber\\
&\mathbf{v} = [\sqrt{\beta_{2}  g_{a b}} \mathbf{h}_{a b}^{H}  \mathbf{v}_{AN}\quad \sqrt{\beta_{2}  g_{a i b}} \mathbf{h}_{i b}^{H}\operatorname{diag}\left(\mathbf{H}_{a i} \mathbf{v}_{AN}\right)].
\end{align}
Similarly, the rate of Eve can be rewritten as
\begin{equation}
R_{e}=\log _{2}\left(1+\frac{P_{t} \left(\overline{\bm{\theta}}^{H} \mathbf{m}^{H} \mathbf{m} \overline{\bm{\theta}}\right)}{P_{t} \left(\overline{\bm{\theta}}^{H} \mathbf{n}^{H} \mathbf{n} \overline{\bm{\theta}}\right)+\sigma_{e}^{2}}\right),
\end{equation}
where
\begin{align}
&\mathbf{m} = [\sqrt{\beta_{1}  g_{a e}} \mathbf{h}_{a e}^{H}  \mathbf{v}_{a} \quad \sqrt{\beta_{1}  g_{a i e}} \mathbf{h}_{i e}^{H}\operatorname{diag}\left(\mathbf{H}_{a i}\mathbf{v}_{a}\right) ],\nonumber\\
&\mathbf{n} = [\sqrt{\beta_{2}  g_{a e}} \mathbf{h}_{a e}^{H}  \mathbf{v}_{AN} \quad \sqrt{\beta_{2}  g_{a i e}} \mathbf{h}_{i e}^{H}\operatorname{diag}\left(\mathbf{H}_{a i}\mathbf{v}_{AN}\right) ].
\end{align}

Therefore, the optimization problem in (\ref{Max-SR}) is converted into
\begin{equation}\label{theta}
\begin{aligned}
\max _{\overline{\bm{\theta}}} &\quad \frac{\overline{\bm{\theta}}^{H} \mathbf{Q} \overline{\bm{\theta}}}{\overline{\bm{\theta}}^{H} \mathbf{K} \overline{\bm{\theta}}} \times \frac{\overline{\bm{\theta}}^{H} \mathbf{T} \overline{\bm{\theta}}}{\overline{\bm{\theta}}^{H} \mathbf{R} \overline{\bm{\theta}}} &\text{s.t.} &\quad \overline{\bm{\theta}}^{H} \overline{\bm{\theta}}=N_{r}+1,\\
\end{aligned}
\end{equation}
where
\begin{align}
&\mathbf{Q}=(P_{t} \mathbf{v}^{H} \mathbf{v}+P_{t} \mathbf{w}^{H} \mathbf{w}+\frac{\mathbf{I}_{N_{r}+1}}{N_{r}+1} \sigma_{b}^{2}),\nonumber\\
&\mathbf{K}=(P_{t} \mathbf{v}^{H} \mathbf{v}+\frac{\mathbf{I}_{N_{r}+1}}{N_{r}+1} \sigma_{b}^{2}),
\mathbf{T}=(P_{t} \mathbf{n}^{H} \mathbf{n}+\frac{\mathbf{I}_{N_{r}+1}}{N_{r}+1} \sigma_{e}^{2}),\nonumber\\
&\mathbf{R}=(P_{t} \mathbf{n}^{H} \mathbf{n}+P_{t} \mathbf{m}^{H} \mathbf{m}+\frac{\mathbf{I}_{N_{r}+1}}{N_{r}+1} \sigma_{e}^{2}).
\end{align}
Finally, the $\overline{\bm{\theta}}$ in (\ref{theta}) can be solved via GPI in \cite{lee2008GPI}. The whole procedure is summarized in the following Table.

\begin{table}[htbp]
 \centering
 \label{tab:pagenum}
 \begin{tabular}{llll}
  \toprule
   \textbf{Algorithm 1} Proposed Max-SR-GPI method\\
  \midrule
  1:   Set initial solution $\bm{\Theta}^{(0)}$, $\mathbf{v}_{a}^{(0)}$ and $\mathbf{v}_{AN}^{(0)}$. Randomly take the value of\\
   $\bm{\Theta}$,
   and calculate the initial $R_{s}^{(0)}$ multiple times based on formula (\ref{SR}).\\
  2:  Set $p$=0, threshold $\epsilon$. \\
  3:  \textbf{repeat}\\
  4: Given ($\bm{\Theta}^{(p)}$,$\mathbf{v}_{AN}^{(p)}$), according to (\ref{va}) to get $\mathbf{v}_{a}^{(p+1)}$.\\
  5: Given ($\bm{\Theta}^{(p)}$,$\mathbf{v}_{a}^{(p+1)}$), according to (\ref{van}) to get $\mathbf{v}_{AN}^{(p+1)}$.\\
  6: Given ($\mathbf{v}_{a}^{(p+1)}$,$\mathbf{v}_{AN}^{(p+1)}$), according to (\ref{theta}) to get $\bm{\Theta}^{(p+1)}$.\\
  7: Compute $R_{s}^{(p+1)}$ using $\mathbf{v}_{a}^{(p+1)}$,$\mathbf{v}_{AN}^{(p+1)}$ and $\bm{\Theta}^{(p+1)}$.\\
  8: $p$=$p$+1;\\
  9: \textbf{until} $R_{s}^{(p)}-R_{s}^{(p-1)} \leq \epsilon$, and record the maximum SR value $R_{s}^{(p)}$.\\
  \bottomrule
 \end{tabular}
\end{table}

The complexity of Max-SR-GPI method is $\mathcal{O}(L_{1}(L_{2}(3(N_{r}+1)^3$$+7(N_{r}+1)^2)+L_{3}(3N_{a}^3$+$7N_{a}^2)+
2N_{a}^3+4N_{a}^2)$ float-point operations (FLOPs), where $L_{1}$, $L_{2}$, and $L_{3}$ denote the iterative numbers of optimization variables $\mathbf{v}_{a}$, $\mathbf{v}_{AN}$, and $\bm{\theta}$.
\subsection{Proposed Max-RP-ZFC}
In the previous subsection, the optimization variables $\mathbf{v}_{AN}$ and $\bm{\theta}$ are computed by the iterative method GPI. The corresponding computational complexity is a linearly increasing function of their numbers of iterations. To reduce the part complexity, a low-complexity method Max-RP-ZFC is proposed to solve $\mathbf{v}_{AN}$ by ZF and $\bm{\theta}$ by  maximizing RP in closed-form.

First, we optimize the AN beamforming vector $\mathbf{v}_{AN}$, which is independent of $\bm{\theta}$ and $\mathbf{v}_{a}$ . Below,  maximize the receive AN power along the direct channel from Alice to Eve at Eve with respect to $\mathbf{v}_{AN}$ is  formualized  as
\begin{subequations}\label{VAN}
\begin{align}
\max _{\mathbf{v}_{AN}}  \quad& \mathbf{v}_{AN}^{H} \mathbf{h}_{ae}\mathbf{h}_{ae}^{H} \mathbf{v}_{AN} \\
\text { s.t. } \quad& \left( \mathbf{h}_{ab} \quad \mathbf{H}_{a i}^{H}\right)^{H} \mathbf{v}_{AN}=\mathbf{0} ,\quad\mathbf{v}_{AN}^{H} \mathbf{v}_{AN}=1.
\end{align}
\end{subequations}
Let us define
$\mathbf{G} = \left( \mathbf{h}_{ab} \quad \mathbf{H}_{a i}^{H}\right)^{H}$, $\mathbf{T}_{-ae}$ = $\left[\mathbf{I}_{N_{a}}-\mathbf{G}^{H}\left(\mathbf{G} \mathbf{G}^{H}\right)^{\dagger} \mathbf{G}\right]$, and $\mathbf{v}_{AN}$=$\mathbf{T}_{-ae}$$\cdot$$\mathbf{u}_{AN}$, then, problem (\ref{VAN}) can be simplified as
\begin{equation}\label{uan}
\begin{aligned}
\max _{\mathbf{u}_{AN}}  \quad& \mathbf{u}_{AN}^{H}\mathbf{T}_{-ae}^{H} \mathbf{h}_{ae}\mathbf{h}_{ae}^{H} \mathbf{T}_{-ae}\mathbf{u}_{AN}
&\text { s.t. } &\mathbf{u}_{AN}^{H} \mathbf{u}_{AN}=1.\\
\end{aligned}
\end{equation}
which directly gives
\begin{equation}
 \mathbf{v}_{AN}={\mathbf{T}_{-ae}  \mathbf{h}_{ae}}/{\left\|\mathbf{T}_{-ae}  \mathbf{h}_{ae}\right\|}
\end{equation}
due to the fact that matrix $\mathbf{T}_{-ae}$ is a rank-one matrix. Now, we establish a joint two-variable ($\mathbf{v}_{a}$ and $\bm{\theta}$) optimization problem of maximizing RP at Bob as follows
\begin{subequations}\label{RP}
\begin{align}
\max _{\mathbf{v}_{a},\bm{\Theta}}\quad & \mathbf{v}_{a}^{H}\left(\mathbf{h}_{ib}^{H} \bm{\Theta} \mathbf{H}_{ai}+\mathbf{h}_{ab}^{H}\right)^{H}\left(\mathbf{h}_{ib}^{H} \bm{\Theta} \mathbf{H}_{ai}+\mathbf{h}_{ab}^{H}\right) \mathbf{v}_{a} \\
\text { s.t. }  \quad &\mathbf{h}^{H}_{ae} \mathbf{v}_{a}=0, \quad\mathbf{v}_{a}^{H} \mathbf{v}_{a}=1, \quad\bm{\theta}^{H}\bm{\theta}=N_{r}.
\end{align}
\end{subequations}
 Similar to (\ref{uan}), fixing $\bm{\Theta}$, we have
\begin{equation}
\mathbf{v}_{a}={\mathbf{P}\left(\mathbf{h}_{ib}^{H} \bm{\Theta}\mathbf{H}_{ai} +\mathbf{h}_{ab}^{H}\right)^{H}}/{\left\| \mathbf{P}\left(\mathbf{h}_{ib}^{H} \bm{\Theta}\mathbf{H}_{ai} +\mathbf{h}_{ab}^{H}\right)^{H}\right\|},
\end{equation}
where $\mathbf{P}$ = $\mathbf{I}_{N_{a}}-\mathbf{h}_{ae} \mathbf{h}_{ae}^{H}$.
Then, fixing $\mathbf{v}_{a}$, (\ref{RP}) can be rewritten as
\begin{subequations}\label{34}
\begin{align}
\max _{\bm{\theta}}\quad &\mathbf{u}_{a}^{H}\mathbf{P}\left(\mathbf{h}_{ib}^{H} \bm{\Theta} \mathbf{H}_{ai}+\mathbf{h}_{ab}^{H}\right)^{H}
\left(\mathbf{h}_{ib}^{H} \bm{\Theta} \mathbf{H}_{ai}+\mathbf{h}_{ab}^{H}\right)\mathbf{P} \mathbf{u}_{a}\\
\text { s.t. } \quad &\bm{\theta}^{H}\bm{\theta}=N_{r},
\end{align}
\end{subequations}
where $\mathbf{u}_{a}^{H}$$\mathbf{u}_{a}$=1. 
 The objective function of (\ref{34}) can be expressed in the following quadratic form
\begin{align}
&\bm{\theta}^{H} \operatorname{diag}\left(\mathbf{u}_{a}^{H}\mathbf{P}^{H}\mathbf{H}_{ai}^{H} \right)  \mathbf{h}_{ib} \mathbf{h}_{i b}^{H}\operatorname{diag}\left(\mathbf{H}_{ai} \mathbf{P} \mathbf{u}_{a}\right)  \bm{\theta}+ \nonumber\\
&\bm{\theta}^{H} \operatorname{diag}\left(\mathbf{u}_{a}^{H}\mathbf{P}^{H}\mathbf{H}_{ai}^{H} \right) \mathbf{h}_{ib} \mathbf{h}_{ab}^{H} \mathbf{P} \mathbf{u}_{a}+ \mathbf{u}_{a}^{H} \mathbf{P}^{H}  \mathbf{h}_{ab} \mathbf{h}_{ib}^{H}\cdot\nonumber\\
&\operatorname{diag}(\mathbf{H}_{ai}\mathbf{P} \mathbf{u}_{a}) \bm{\theta}+\mathbf{u}_{a}^{H} \mathbf{P}^{H}  \mathbf{h}_{ab} \mathbf{h}_{ab}^{H} \mathbf{P} \mathbf{u}_{a}.
\end{align}
Substituting the above expression in (\ref{34}) yields

\begin{align}\label{QCEC}
&\max_{\bm{\theta}} ~\bm{\theta}^{H} \mathbf{A} \bm{\theta}+\bm{\theta}^{H} \mathbf{b}+\mathbf{b}^{H} \bm{\theta}+C
&\text { s.t. }  ~\bm{\theta}^{H} \bm{\theta}=N_{r},
\end{align}
where
\begin{align}
&\mathbf{A}=\underbrace{\operatorname{diag}\left(\mathbf{u}_{a}^{H}\mathbf{P}^{H}\mathbf{H}_{ai}^{H}\right)\mathbf{h}_{ib}}_{\mathbf{a}}\underbrace{\mathbf{h}_{i b}^{H}\operatorname{diag}\left(\mathbf{H}_{ai} \mathbf{P} \mathbf{u}_{a}\right)}_{\mathbf{a}^H},\nonumber\\
&\mathbf{b}=\operatorname{diag}\left(\mathbf{u}_{a}^{H}\mathbf{P}^{H}\mathbf{H}_{ai}^{H} \right) \mathbf{h}_{ib} \mathbf{h}_{ab}^{H} \mathbf{P} \mathbf{u}_{a}, \nonumber\\
&C=\mathbf{u}_{a}^{H} \mathbf{P}^{H}  \mathbf{h}_{ab} \mathbf{h}_{ab}^{H} \mathbf{P} \mathbf{u}_{a}.
\end{align}
The Lagrangian function of (\ref{QCEC}) can be expressed as
\begin{align}\label{aim}
f(\bm{\theta}, \lambda)=\bm{\theta}^{H} \mathbf{A} \bm{\theta}+\bm{\theta}^{H} \mathbf{b}+\mathbf{b}^{H} \bm{\theta}+C+\lambda\left(\bm{\theta}^{H} \bm{\theta}-N_{r}\right).
\end{align}
whose partial derivative with respect to $\bm{\theta}$ is set to $0$ to obtain the following equation
\begin{equation}
\frac{\partial f(\bm{\theta}, \lambda)}{\partial \bm{\theta}}=\mathbf{A} \bm{\theta}+\lambda \bm{\theta}+\mathbf{b}=0.
\end{equation}
which generates
\begin{equation}
\bm{\theta}=-(\mathbf{A}+\lambda \mathbf{I}_{N_{r}})^{-1}\mathbf{b}.
\end{equation}
Since $\mathbf{A}=\mathbf{a}\mathbf{a}^H$ is a matrix of rank-one, using the Sherman-Morrison formula,
the constraint of (\ref{QCEC}) can be expressed as
\begin{align}\label{s.t.}
\mathbf{b}^{H}\left(\frac{\mathbf{I}_{N_{r}}}{\lambda^{2}}+\frac{\left(\mathbf{a} \mathbf{a}^{H}\right)^{2}-2 \mathbf{a} \mathbf{a}^{H}}{\lambda^{2}\left(\lambda+\mathbf{a}^{H} \mathbf{a}\right)^{2}}\right) \mathbf{b}=N_{r},
\end{align}
which can be simplified as
\begin{equation}\label{lambda}
\begin{aligned}
&N_{r} \lambda^{4}+2 N_{r} a_{1} \lambda^{3}+\left(N_{r} a_{1}^{2}-b_{1}\right) \lambda^{2}- \\
&2 c_{1} \lambda-a_{1}^{2} b_{1}-d_{1}+2 e_{1}=0,
\end{aligned}
\end{equation}
where
\begin{align}
&a_{1}=\mathbf{a}^{H}\mathbf{a}, b_{1}=\mathbf{b}^{H}\mathbf{b}, c_{1}=\mathbf{b}^{H}\mathbf{b}\mathbf{a}^{H}\mathbf{a},\nonumber\\
&d_{1}=\mathbf{b}^{H}\mathbf{a}\mathbf{a}^{H}\mathbf{a}\mathbf{a}^{H}\mathbf{b}, e_{1}=\mathbf{b}^{H}\mathbf{a}\mathbf{a}^{H}\mathbf{b}.
\end{align}
Observing equation (\ref{lambda}), it is a fourth-order polynomial  and has a set $S$ of four roots denoted as: $\lambda_1$, $\lambda_2$, $\lambda_3$, and $\lambda_4$.   Finding the optimal value of $\lambda_{0}$ is modelled as the following maximum problem
\begin{align}
\lambda_{0}=\underset{\lambda \in\{S\}}{\arg \max }  ~~(\ref{aim}).
\end{align}

Alternating iteration between $\mathbf{v}_{a}$ and $\bm{\theta}_{i}$ are repeated until $R_{s}^{(p)}-R_{s}^{(p-1)} \leq \epsilon$.

The complexity of Max-RP-ZFC is $\mathcal{O}(L_{4}(N_{r}^3+7N_{r}^2+2N_{r}^2N_{a}+$$14N_{a}^2+6N_{a}^2N_{r}
   -4N_{r}N_{a}-6N_{a}-2N_{r})+2N_{a}^2+N_{a})$ FLOPs, where $L_{4}$ denotes the alternating iterative number between $\mathbf{v}_{a}$ and $\bm{\theta}$.

\section{Simulation Results and Discussions}
In this section,  numerical simulation results are presented to evaluate the SR and convergent performance of our proposed methods. Simulation parameters are set as follows: $P_{s}$ = 30 dBm, $\sigma_{b}^{2}=\sigma_{e}^{2}$=-40dBm, and $N_{a}$ = 16. The distances and angles are set as $d_{ai}$ = 20 m, $d_{ab}$ = 40 m,  $d_{ae}$ = 50 m,  $\theta_{ai}$ = $29\pi/60$, $\theta_{ab}$ = $1\pi/2$, and $\theta_{ae}$ = $23\pi/36$.

Fig.~\ref{diedai.eps} shows the SR versus the number of iterations for three typical number of elements of IRS as follows: 32, 128, and 1024. From Fig.2, it is very clear that  the proposed two methods rapidly converge to the SR ceil with only $3\sim 5$ iterations. Also, we find that the SR performance gain achieved by IRS is very attractive as the number of element of IRS increases from small-scale to large-scale.
\begin{figure}[htbp]
\centering
\includegraphics[width=0.4\textwidth]{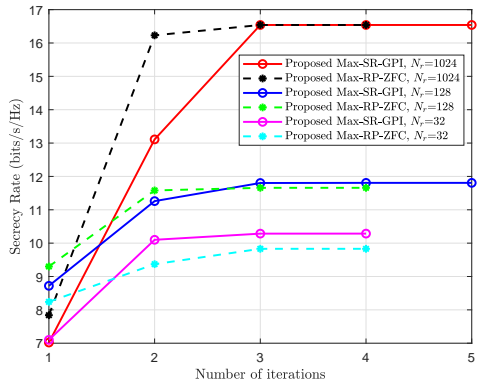}\\
\caption{Convergent curves of proposed algorithms at different numbers of IRS elements}\label{diedai.eps}
\end{figure}

\begin{figure}[htbp]
\centering
\includegraphics[width=0.40\textwidth]{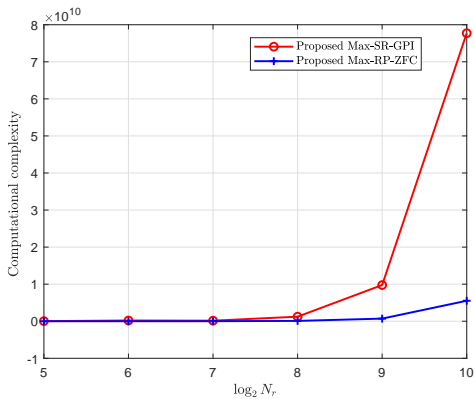}\\
\caption{Computational complexity versus the number of IRS elements}\label{FLOP.eps}
\end{figure}
 Fig.~\ref{FLOP.eps} depicts the curves of computational complexity versus the number of IRS elements. For small-scale or medium-scale IRS, the proposed two methods have the same complexity. Conversely,  for large-scale IRS, the complexity of the proposed Max-SR-GPI is far higher than that of Max-RP-ZFC.

\begin{figure}[htbp]
\centering
\includegraphics[width=0.40\textwidth]{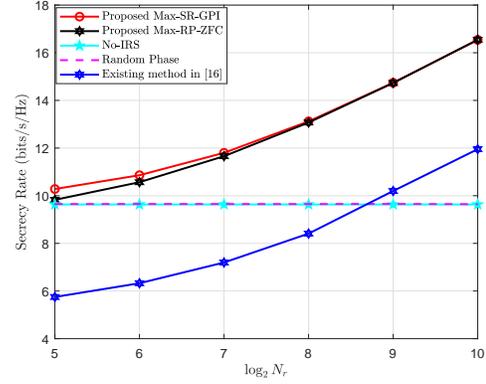}\\
\caption{Secrecy rate versus the number of IRS elements}\label{RS_Nr.eps}
\end{figure}
Fig.~\ref{RS_Nr.eps} plots the SR versus the number of IRS elements $N_{r}$ for our proposed two methods  with no IRS and random phase as performance benchmarks. The SR performance of the proposed two methods is much better than  those of no IRS, random phase and existing methods and gradually grow with $N_r$.  The proposed Max-SR-GPI performs better than the Max-RP-ZFC in accordance with SR  when $N_{r}$ is small-scale.

%

\section{Conclusions}
In this paper, we have made an investigation of the  IRS-assisted DM networks with single-CM-stream transmission. In order to improve the SR performance, two high-performance methods Max-SR-GPI method and Max-RP-ZFC  were proposed. Simulation results showed that the proposed two methods harvest obvious SR performance gains over no IRS, random phase, and existing method \cite{Hujingsong2020}, especially in the case of large-scale IRS. Additionally, they can converge rapidly. The  Max-RP-ZFC is far lower than Max-SR-GPI in terms of computational complexity, particularly in the case of large-scale IRS.
\ifCLASSOPTIONcaptionsoff
\newpage
\fi

\bibliographystyle{IEEEtran}
\bibliography{IEEEfull,reference}
\end{document}